\documentclass{pic2012}
\def\runauthor{L.~J.~Kaufman}
\def\shorttitle{Double Beta Decay}
\begin{document}

\title{Recent Results in Neutrinoless Double Beta Decay 
}

\author{Lisa J. Kaufman}

\address{Indiana University Department of Physics and CEEM\\
Bloomington, IN 47405, USA\\
E-mail: ljkauf@indiana.edu }

\maketitle

\abstracts{The search for neutrinoless double beta decay is a rich source for new physics. The observation of this decay will lead to understanding of the absolute mass scale of neutrinos, the Majorana nature of the neutrino (whether the neutrino is its own anti-particle), and lepton number violation. Double beta decay is being investigated around the world by several experiments using different candidate isotopes. There has been much progress made in experimental techniques recently such that achieving sensitivity to neutrino masses at 50 meV and below will be possible in the near future. A summary of recent results in neutrinoless double beta decay is discussed with a look toward the experimental goals for the future.
}

\section{Introduction} 
The Standard Model double beta decay ($2\nu\beta\beta$) process $N \rightarrow N^{\prime}e^-e^- \bar{\nu}_e \bar{\nu}_e$ is a second-order weak process that can occur for even-even nuclides where $\beta$ decay is either energetically forbidden or suppressed due to large angular momemtum differences.  The $2\nu\beta\beta$ decay has been observed in many isotopes~\cite{RPP}, first in $^{82}$Se in 1987 and most recently in $^{136}$Xe with the EXO-200 detector~\cite{EXO200_2nu}, later confirmed in~\cite{KLZ_2nu}.

The hypothetical neutrinoless ($0\nu\beta\beta$) decay process process $N \rightarrow N^{\prime}e^- e^-$ is the best tool we have for addressing the Majorana/Dirac nature of neutrinos.  This decay can only occur if neutrinos are massive and do not carry lepton number; therefore, it can only occur for Majorana neutrinos~\cite{schechter_valle}.  The recent discovery of neutrino mass from oscillation experiments~\cite{RPP} makes the search for the Majorana nature of neutrinos particularly relevant.  In addition, assuming the $0\nu\beta\beta$ process occurs due to the exchange of light Majorana neutrinos, the decay rate is directly proportional to the square of an effective Majorana neutrino mass, $\langle m_{\beta\beta} \rangle$ (i.e. a coherent superposition of the electron neutrino projections on the mass eigenstates),  by the product of phase space and a nuclear matrix element squared.  As a result, measurement of neutrinoless double beta decay not only tells us about the Majorana/Dirac nature of the neutrino but also the absolute mass scale of the neutrino.  Kinematic measurements restrict the neutrino mass scale to be below $\mathcal{O}(1\ {\rm eV})$~\cite{kine}, leading to $0\nu\beta\beta$ half-lives beyond $10^{24}$\,yr. Nuclei that undergo two-neutrino double beta decay are candidates for neutrinoless double beta decay.  An active research program worldwide using these nuclei to search for $0\nu\beta\beta$ decay is underway.  A search for $0\nu\beta\beta$ decay in $^{76}$Ge has claimed a positive observation~\cite{Klapdor} with $T^{0\nu\beta\beta}_{1/2}$($^{76}$Ge)$ = (2.23^{+0.44}_{-0.31})\times 10^{25}$\,yr, implying $\langle m_{\beta\beta} \rangle = 0.32\pm 0.03$\,eV for the nuclear matrix element given in \cite{QRPA2}.

In order to probe the allowed region for the inverted neutrino mass hierarchy, experiments will have to be sensitive to half-lives of the order 10$^{27}$ years or longer which corresponds to neutrino  $\langle m_{\beta\beta} \rangle \leq 50$ meV depending on the nuclear matrix elements.  Current experiments are sensitive to half-lives of 10$^{25}$ years and  $\langle m_{\beta\beta} \rangle \simeq 100$ meV.  Because of the uncertainty in the nuclear matrix element calculations~\cite{QRPA2,GCM,NSM,IBM-2,QRPA1}  used to determine $\langle m_{\beta\beta} \rangle$ and different gamma backgrounds for various isotopes, it is important to have a double beta program that uses several different isotopes to measure the $0\nu\beta\beta$ half-life~\cite{GehmanElliott}.

\section{Experimental Technique}

The exceedingly long half-lives of interest for $0\nu\beta\beta$ decay require large detectors using isotopically enriched sources, radio-clean construction techniques and the ability to actively reject backgrounds due to various types of radioactivity.  The $0\nu\beta\beta$ decay results in a discrete electron sum energy distribution centered at the Q-value ($Q_{\beta\beta}$).  The $2\nu\beta\beta$ decay is characterized by a continuous sum energy spectrum ending at $Q_{\beta\beta}$.  A high resolution measurement of the decay electron sum energy allows discrimination between the two decay modes.

There are currently three double beta decay experiments actively taking data: EXO-200, KamLAND-Zen, and \textsc{Gerda}.  EXO-200 and KamLAND-Zen are searching for $0\nu\beta\beta$ in $^{136}$Xe while \textsc{Gerda} is searching the decay in the $^{76}$Ge nucleus.  Each experiment follows a similar experimental design: minimize cosmogenic backgrounds by placing the experiment deep underground, put a large mass of the isotope of interest in a detector at the center of a shielded volume to minimize backgrounds from high energy gamma rays that come from the uranium and thorium content of the rock underground, build the detector out of radio-pure material to minimize gamma backgrounds from internal components, collect ionization (and scintillation) signal for best energy resolution possible and segment detector for position resolution.  More details regarding the individual experiments will be discussed in the following sections.

Several other double beta decay experiments are currently in the construction phase and will begin taking data soon.  The CUORE experiment at Gran Sasso is a bolometric experiment using $^{130}$Te which is naturally 34\% abundant.  The \textsc {Majorana Demonstrator} at the Sanford Underground Laboratory is a germanium semiconductor using germanium enriched to 86\% with $^{76}$Ge.  SNO+ at SNOLab is a liquid scintillator detector currently designed to use $^{150}$Nd with a natural abundance of 5.6\%.  Other experiments being in various stages of research and development SuperNEMO ($^{82}$Se, source foil tracking and scintillation), NEXT ($^{136}$Xe, gas time projection chamber), COBRA ($^{116}$Cd, CdZnTe semiconductor), MOON ($^{100}$Mo, source foil and plastic scintillator) and CANDLES ($^{48}$Ca, CaF$_2$ scintillator).

\subsection{EXO-200}

EXO-200, described in detail in~\cite{EXO200_det1}, uses xenon both as source and detector for the two electrons emitted in its $\beta\beta$ decay. The detector is a cylindrical homogeneous time projection chamber (TPC)~\cite{TPC}.  It is filled with liquefied xenon ($^{enr}$LXe) enriched to (80.6$\pm$0.1)\% in the isotope $^{136}$Xe. The remaining 19.4\% is $^{134}$Xe, with other isotopes present only at low concentration. EXO-200 is designed to minimize radioactive backgrounds, maximize the $^{enr}$LXe fiducial volume, and provide good energy resolution at the $^{136}$Xe Q-value of $2457.83\pm0.37$ keV~\cite{Redshaw}. Energy depositions in the TPC produce both ionization and scintillation signals. The TPC configuration allows for three-dimensional topological and temporal reconstruction of individual energy depositions. This ability is essential for discriminating $\beta\beta$ decays from residual backgrounds dominated by $\gamma$s. 

\begin{figure}
\begin{center}
	\resizebox{7cm}{!}{\includegraphics{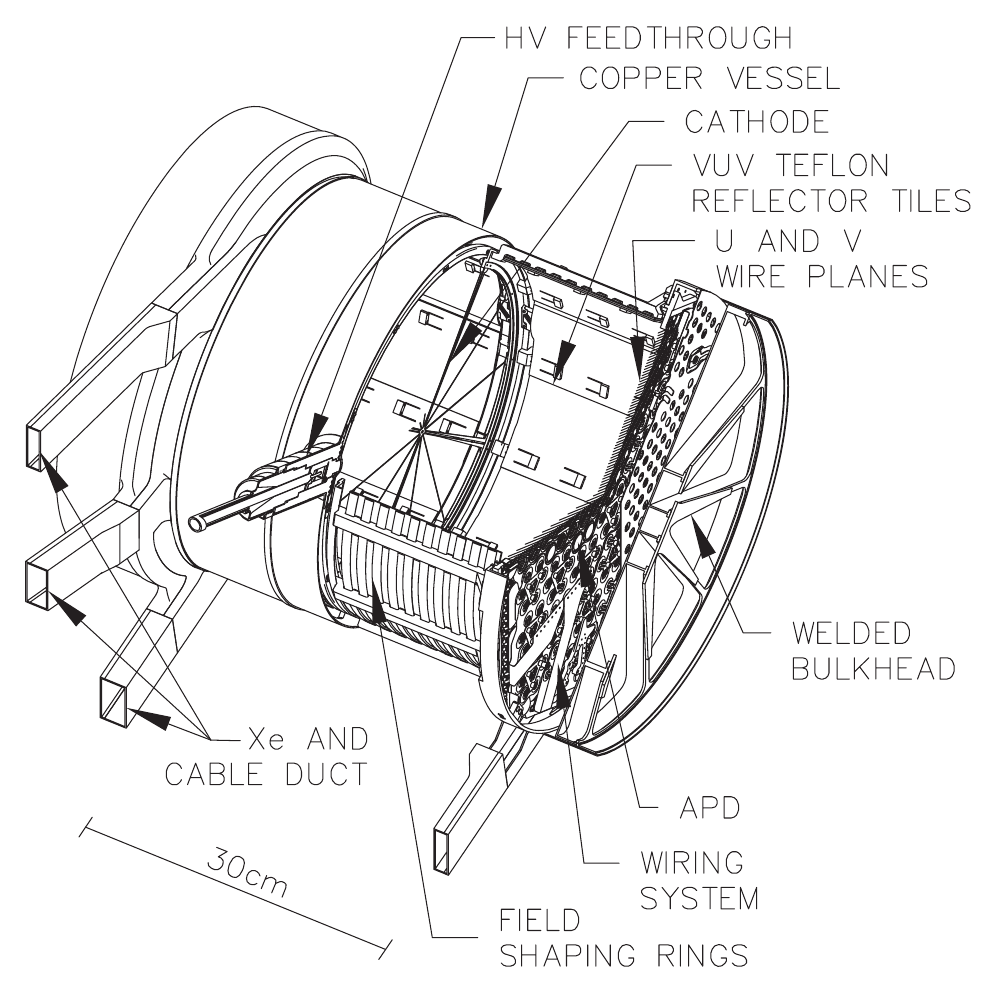}}
	\caption[EXO-200 TPC]{The cylindrical EXO-200 TPC. Each end is intrumented for scintillation light and stereoscopic charge readout. See text for details.}
	\label{fig:tpc}
\end{center}
\end{figure}

The cylindrical TPC (Figure~\ref{fig:tpc}) is divided into two symmetric volumes separated by a cathode grid. Each end of the TPC is instrumented with charge induction (V) and charge collection (U) wires with a pitch of 9~mm.  The U and V wire grids, crossing at $60^{\circ}$, provide stereoscopic information on the topology of the charge deposition.  At each end of the TPC there is a plane of Large Area Avalanche Photodiodes (LAAPDs) \cite{APDs} that record the 178\,nm scintillation light.  All detector components were carefully selected to minimize internal radioactivity~\cite{activity}. The TPC is mounted in the center of a low-background cryostat.  At least 50\,cm of high purity HFE-7000 fluid~\cite{HFE} and at least 25\,cm of lead shield the TPC from external radioactivity.  The $^{enr}$LXe is continuously purified in gas phase by recirculation~\cite{pump} through hot Zr getters~\cite{SAES}.  
 A calibration system allows the insertion of radioactive sources to various positions immediately outside of the TPC. The clean room module housing the TPC is surrounded on four sides by an array of plastic scintillator panels, serving as a cosmic ray veto.  EXO-200 is located at a depth of $(1585^{+11}_{-6})$ m.w.e. at the Waste Isolation Pilot Plant (WIPP), near Carlsbad, New Mexico, USA. 

EXO-200 started taking low background data in late May 2011.  Data collected until August 2011 was used to measure the $2\nu\beta\beta$-decay rate of $^{136}$Xe \cite{EXO200_2nu}.  The most recent data used for the current $0\nu\beta\beta$ result was collected from September 22, 2011 to April 15, 2012 with a livetime of 120 days and 32.5 kg$\,\cdot\,$yr exposure.

\begin{figure}
\begin{center}
	\resizebox{8cm}{!}{\includegraphics{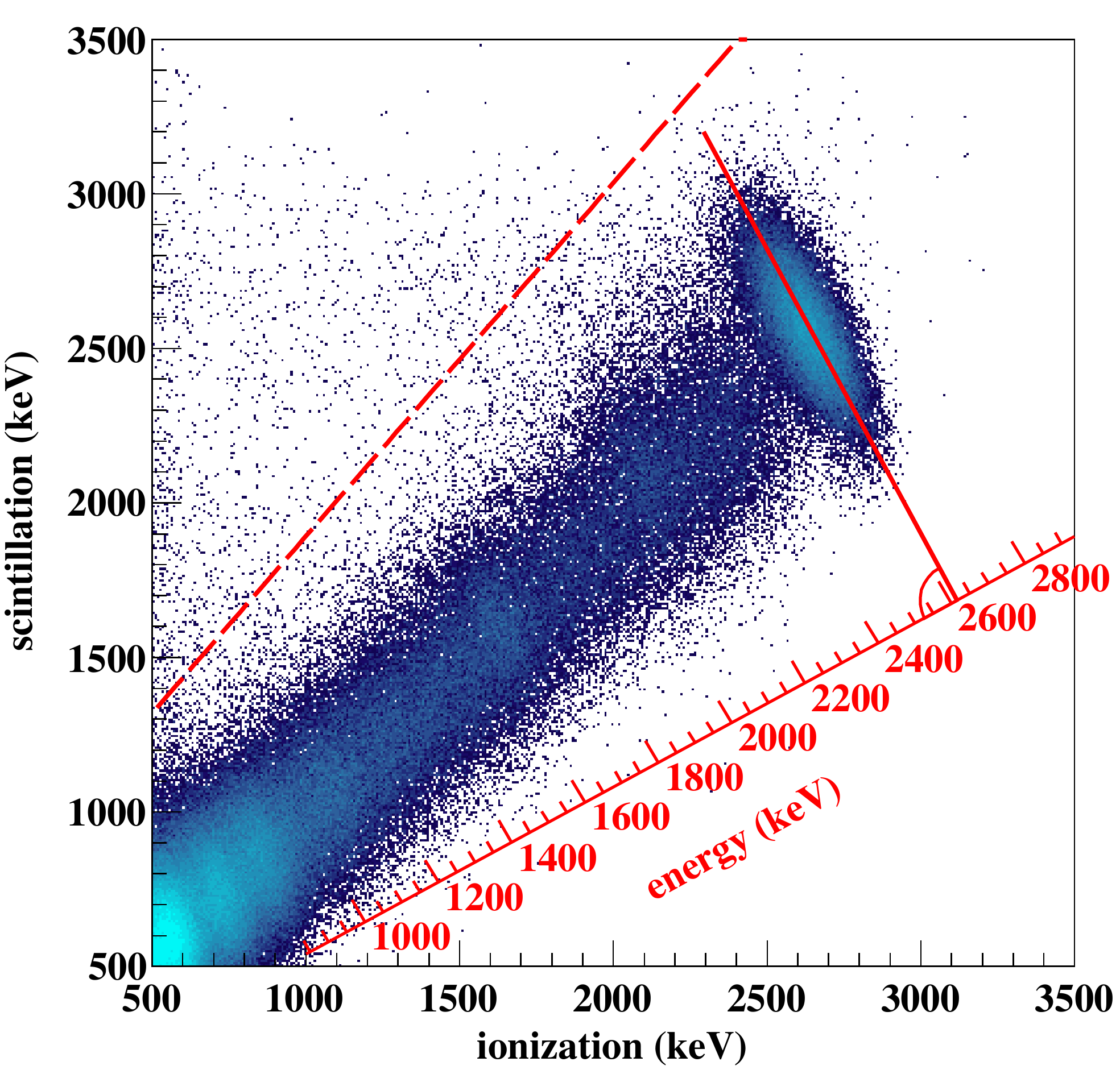}}
	\caption{Correlation between ionization and scintillation for SS events from a $^{228}$Th source. 
Events in the top-left quadrant are due to incomplete charge collection and are rejected by the cut (dashed line), removing only 0.5\% of the total.  The cut is defined using the gamma ray full absorption islands from three calibration sources.}
	\label{fig:correlation}
\end{center}
\end{figure}


Figure~\ref{fig:correlation} shows the energy of events as measured by the ionization and scintillation channels while the $^{228}$Th source was deployed.  As first discussed in~\cite{Conti_etal} and evident from the tilt of the 2615\,keV full absorption ellipse in the figure, the magnitude of the two signals is anticorrelated.  The 2D single-site (SS) and multiple-site (MS) energy spectra are independently rotated and projected onto a new (1D) energy variable in such a way as to minimize the width of the 2615\,keV $\gamma$ line.  Energy spectra from $^{137}$Cs, $^{60}$Co, and $^{228}$Th sources are produced using this method and then, the positions of the full absorption peaks at 662, 1173, 1333 and 2615\,keV are fit to provide the energy calibration.  The ability of the TPC to identify SS and MS interactions is used to separate $\beta$ and $\beta\beta$ decays in the bulk xenon from multiple site $\gamma$ interactions. The clustering, currently applied in 2D, has a separation resolution of 18\,mm in the U-dimension and 6\,mm in z (drift time).

\begin{figure}[t]
\begin{center}
	\resizebox{8cm}{!}{\includegraphics{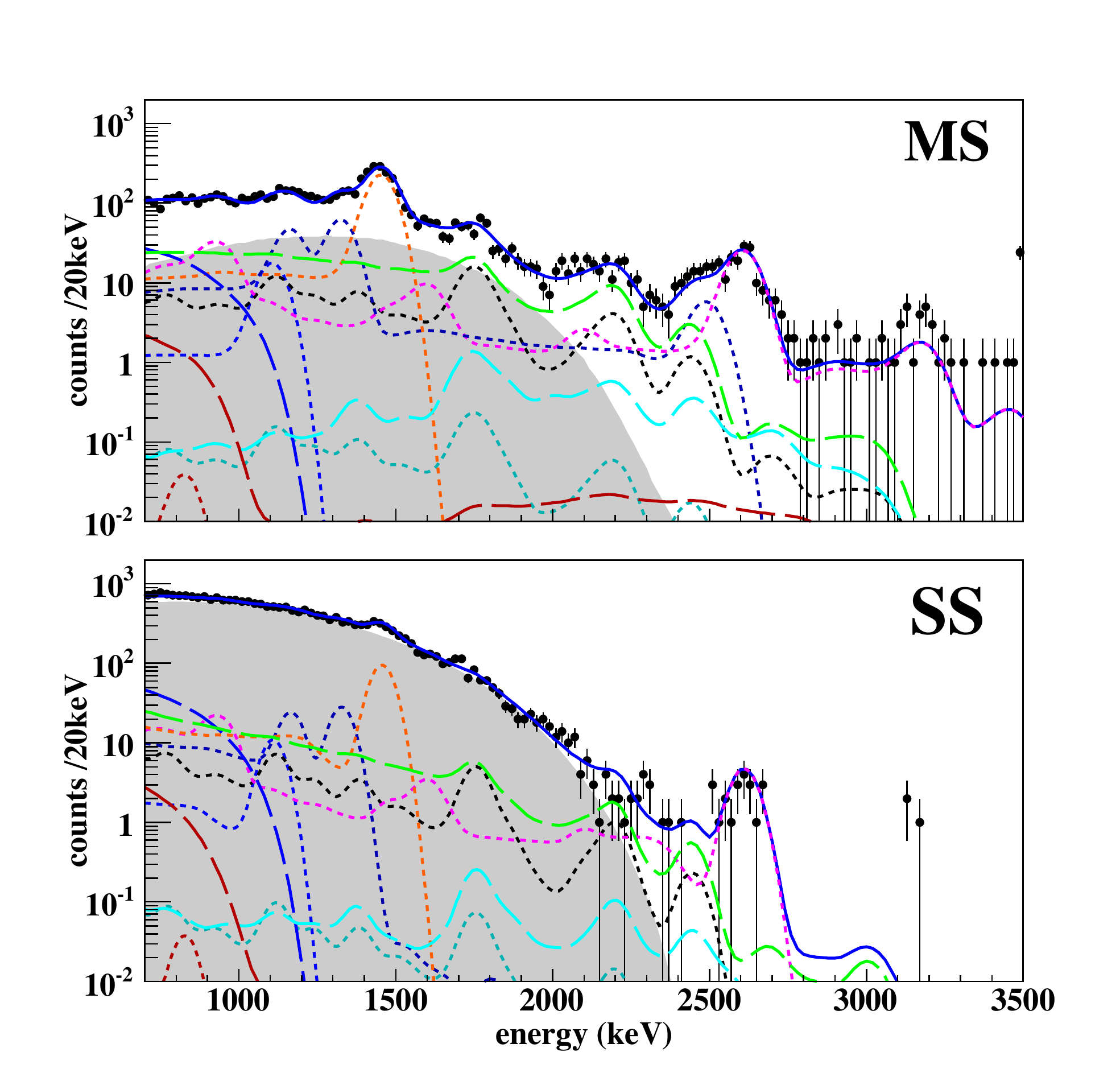}}
	\caption{EXO-200 MS (top) and SS (bottom) energy spectra from the low background data used for the $0\nu\beta\beta$ analysis.  The best fit line (solid blue) is shown.  The background components are $2\nu\beta\beta$ (grey region), $^{40}$K (dotted orange), $^{60}$Co (dotted dark blue), $^{222}$Rn in the cryostat-lead air-gap (long-dashed green), $^{238}$U in the TPC vessel (dotted black), $^{232}$Th in the TPC vessel (dotted magenta), $^{214}$Bi on the cathode (long-dashed cyan), $^{222}$Rn outside of the field cage (dotted dark cyan), $^{222}$Rn in active xenon (long-dashed brown), $^{135}$Xe (long-dashed blue) and $^{54}$Mn (dotted brown). The last bin on the right includes overflows.  There are no overflows in the SS spectrum.}
	\label{fig:low-back}
\end{center}
\end{figure}

\begin{figure}[t]
\begin{center}
	\resizebox{8cm}{!}{\includegraphics{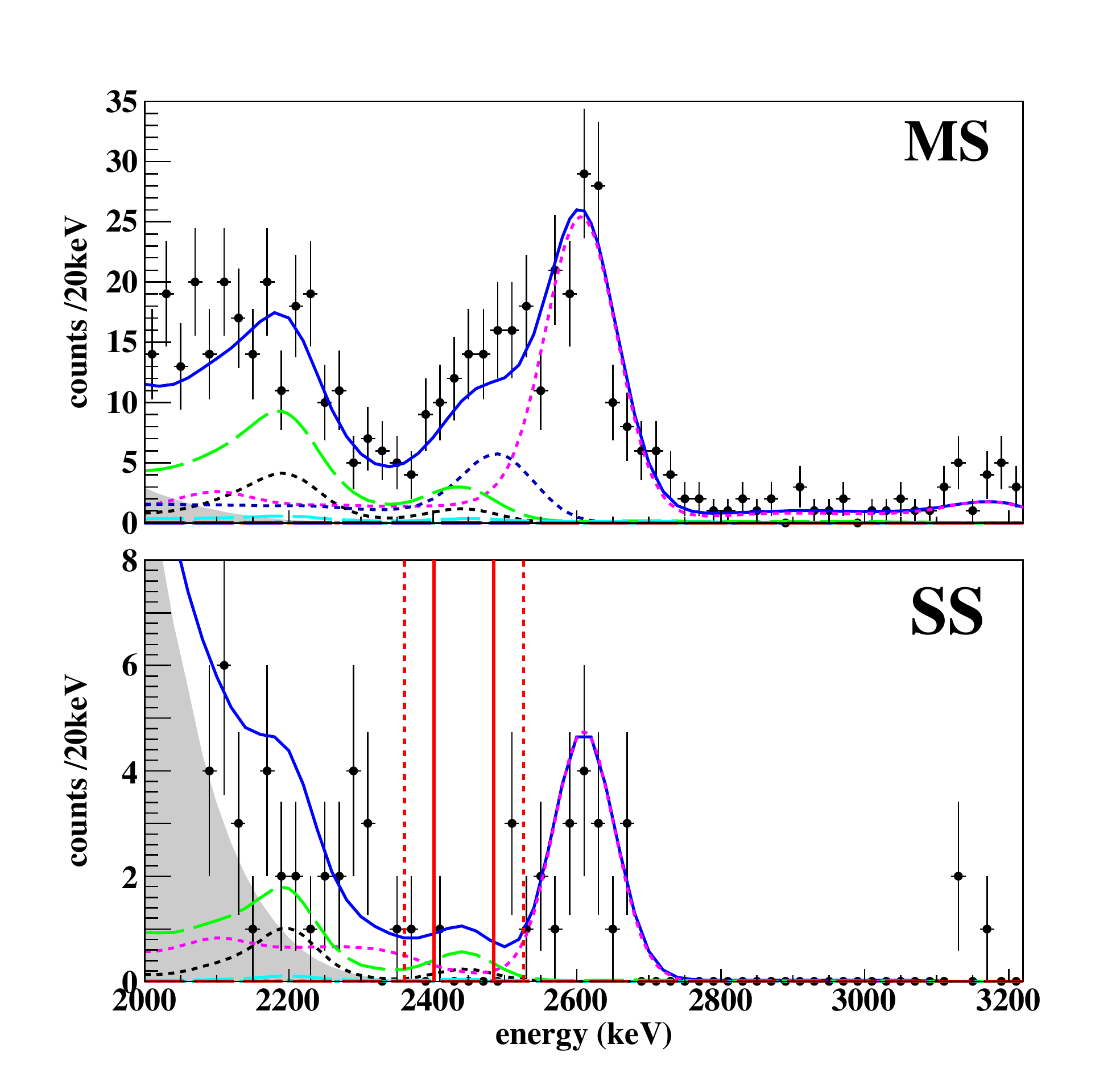}}
	\caption{EXO-200 energy spectra in the $^{136}$Xe $Q_{\beta\beta}$ region for MS (top) and SS (bottom) events.  The 1 (2)$\sigma$ regions around $Q_{\beta\beta}$ are shown by solid (dashed) vertical lines.  The $0\nu\beta\beta$ PDF from the fit is not visible.  The fit results have the same meaning as in Figure~\ref{fig:low-back}.}
	\label{fig:blip}
\end{center}
\end{figure}

The SS and MS low background spectra are shown in Figure~\ref{fig:low-back}.  Primarily due to bremsstrahlung, a fraction of $\beta\beta$ events are MS.  Using a maximum likelihood estimator, the SS and MS spectra are simultaneously fit with PDFs of the $2\nu\beta\beta$ and $0\nu\beta\beta$ of $^{136}$Xe along with PDFs of various backgrounds.  Background models were developed for various components of the detector.  Results of the material screen campaign, conducted during construction, provide the normalization for the models.  The contributions of the various background components to the $0\nu\beta\beta$ and $2\nu\beta\beta$ signal regions were estimated using a previous generation of the detector simulation~\cite{EXO200_det1}.  For the current exposure, the background model treats the activity of the $^{222}$Rn in the air-gap between the cryostat and the lead shielding as a surrogate for all $^{238}$U-like activities external to the cryostat, because of their degenerate spectral shapes and/or small contributions.

For the best-fit energy scale and resolution the $\pm 1\sigma$ and $\pm 2 \sigma$ regions around $Q_{\beta\beta}$ are shown in Figure~\ref{fig:blip}.   The number of events observed in the SS spectrum are 1 and 5, respectively, with the 5 events in the $\pm 2\sigma$ region accumulating at both edges of the interval.  Therefore, no evidence for $0\nu\beta\beta$ decay is found in the present data set.  The upper limit on $T^{0\nu\beta\beta}_{1/2}$ is obtained by the profile likelihood fit to the entire SS and MS spectra.  Systematic uncertainties are incorporated as constrained nuisance parameters.  The fit yields 
an estimate of $4.1\pm0.3$ background counts in the $\pm 1\sigma$ region, giving an expected background rate of $(1.5\pm0.1)\times 10^{-3}$\,kg$^{-1}$yr$^{-1}$keV$^{-1}$.
The fit  also reports $0\nu\beta\beta$ decay limits of $<2.8$ counts at 90\% CL ($<1.1$ at 68\% CL).  This fit corresponds to a $T^{0\nu\beta\beta}_{1/2} > 1.6\times 10^{25}$\,yr at 90\% CL ($T^{0\nu\beta\beta}_{1/2} > 4.6\times 10^{25}$\,yr at 68\% CL).  This result provides the most stringent limit on $\langle m_{\beta\beta} \rangle \leq 140-380$ meV (at the 90\% CL) for the range in nuclear matrix element calculations.  The same fit also reports $T_{1/2}^{2\nu\beta\beta} =(2.23\pm 0.017~{\rm stat.}\pm 0.22~{\rm syst.})\times 10 ^{21}$\,yr, in agreement with~\cite{EXO200_2nu} and \cite{KLZ_2nu}.  This result is in tension with the current observation claim~\cite{Klapdor} as seen in Figure~\ref{fig:comparison}.  More details can be found in~\cite{EXO200_0nu}.

\subsection{KamLAND-Zen}

\begin{figure}
\begin{center}
  \resizebox{8cm}{!}{\includegraphics{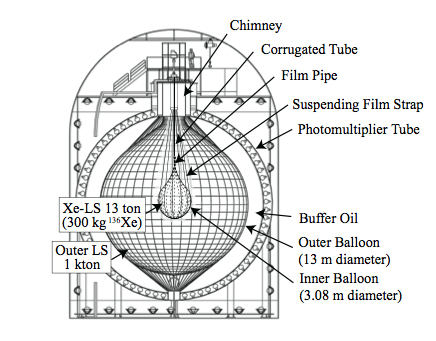}}
  \caption[KamLAND-Zen detector schematic.]{Schematic diagram of the KamLAND-Zen detector (Figure from~\cite{KLZ_2nu}). } 
    \label{fig:KLZ_det}
\end{center}
\end{figure}

KamLAND-Zen (Figure~\ref{fig:KLZ_det}) uses 13 tons of Xe-loaded liquid scintillator (Xe-LS) as the $\beta\beta$ source and detector enriched to 90\% $^{136}$Xe.  The Xe-LS is contained in a 3.08 m diameter nylon inner balloon (IB) suspended at the center of the KamLAND detector.  The KamLAND detector is described in detail in~\cite{KLZ_det}.  The IB is surrounded by liquid scintillator (LS) in an outer balloon that serves as an active shield for external gammas and as a detector for internal radiation from the IB and Xe-LS.  The outer balloon is surrounded by an 18-m spherical stainless steel tank that contains buffer oil to shield the LS from external radiation.  The scintillation light is detected by PMTs that are mounted on the stainless steel tank.

The results from KamLAND-Zen data obtained between October 12, 2011 and February 9, 2012 will be discussed here~\cite{KLZ_majoron}.  The summed energy is determined from the scintillation light produced by the two coincident decay electrons.  The energy response of the detector is calibrated by gammas from a $^{208}$Tl (ThO$_2$W) source, betas and gammas from $^{214}$Bi from a $^{222}$Rn source and 2.225 MeV gammas from spallation neutrons that capture on protons.  The ThO$_2$W source is contained in a lead capsule and deployed close to the outer surface of the IB for calibration, and the most intense peak is produced at 2.6 MeV, close to the Q$_{\beta\beta}$ of $^{136}$Xe.  The energy resolution is determined at 2.6 MeV to be (6.6 $\pm$ 0.3)\%.  

There are three types of backgrounds for the KamLAND-Zen experiment.  The first of these backgrounds is external to the Xe-LS mainly from the IB.  These IB surface events are dominated by $^{134}$Cs in the $2\nu\beta\beta$ energy window and by $^{214}$Bi in the $0\nu\beta\beta$ energy window.  The activity of $^{134}$Cs compared to $^{137}$Cs is consistent with contamination from fallout of the Fukushima reactor accident in March 2011~\cite{KLZ_2nu}.  The second category of backgrounds are uranium and thorium that are internal to the Xe-LS, and the concentrations are estimated by measuring the rate of sequential decays of $^{214}$Bi-$^{214}$Po and $^{212}$Bi-$^{212}$Po.  The 2.6 MeV $\gamma$ from $^{208}$Tl is not a serious background to the $0\nu\beta\beta$ decay because the LS serves as an active veto by detecting the coincident $\beta$ and $\gamma$.  The third type of background is from spallation neutrons that are detected by coincidence of neutron capture gammas with preceding muons, and the events in the energy ranges of interest are small at $<$~0.3~(ton$\cdot$day)$^{-1}$.

\begin{figure}
\begin{center}
  \resizebox{8cm}{!}{\includegraphics{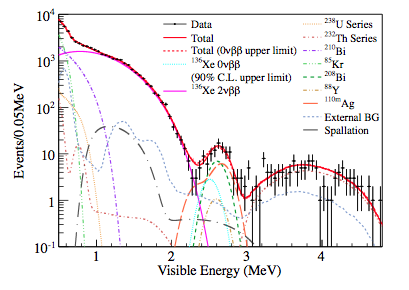}}
  \caption[KamLAND-Zen $0\nu\beta\beta$ energy spectrum fit.]{KamLAND-Zen energy spectrum of selected $\beta\beta$ decay candidates together with the best-fit backgrounds and $2\nu\beta\beta$, and the 90\% CL upper limit for $0\nu\beta\beta$ decays (Figure from~\cite{KLZ_2nu}). } 
    \label{fig:KLZ_fit}
\end{center}
\end{figure}

The data were taken witha live time of  112.3 days and and exposure of 38.6 kg$\cdot$yr of $^{136}$Xe.  The $^{136}$Xe $2\nu\beta\beta$ and $0\nu\beta\beta$ decay rates are estimated from a likelihood fit to the binned ergy spectrum of the selected events between 0.5 MeV and 4.8 MeV.  The best fit spectral decomposition is shown in Figure~\ref{fig:KLZ_fit}.  From these data the $2\nu\beta\beta$ half-life is measured to be $T_{1/2}^{2\nu} = 2.30 \pm 0.02 (\textrm{stat}) \pm 0.12(\textrm{syst}) \times 10^{21}$ yr~\cite{KLZ_majoron}.  A lower limit to the half-life for $0\nu\beta\beta$ is improved from the first result~\cite{KLZ_2nu} to be $T_{1/2}^{0\nu} \geq 6.2 \times 10^{24}$ yr at 90\% CL~\cite{KLZ_majoron}.  This half-life limit corresponds to an upper limit on $\langle m_{\beta\beta} \rangle \leq  260-540$~meV at the 90\% CL.  There is a large $^{110m}$Ag background found in the $0\nu\beta\beta$ energy window potentially that may be from fallout from Fukushima based on ex-situ measurements of the soil around Fukushima.  KamLAND-Zen is undergoing a campaign to remove contaminants from the Xe-LS in order to reduce these backgrounds.

\subsection{GERDA}

\begin{figure}
\begin{center}
  \resizebox{8cm}{!}{\includegraphics{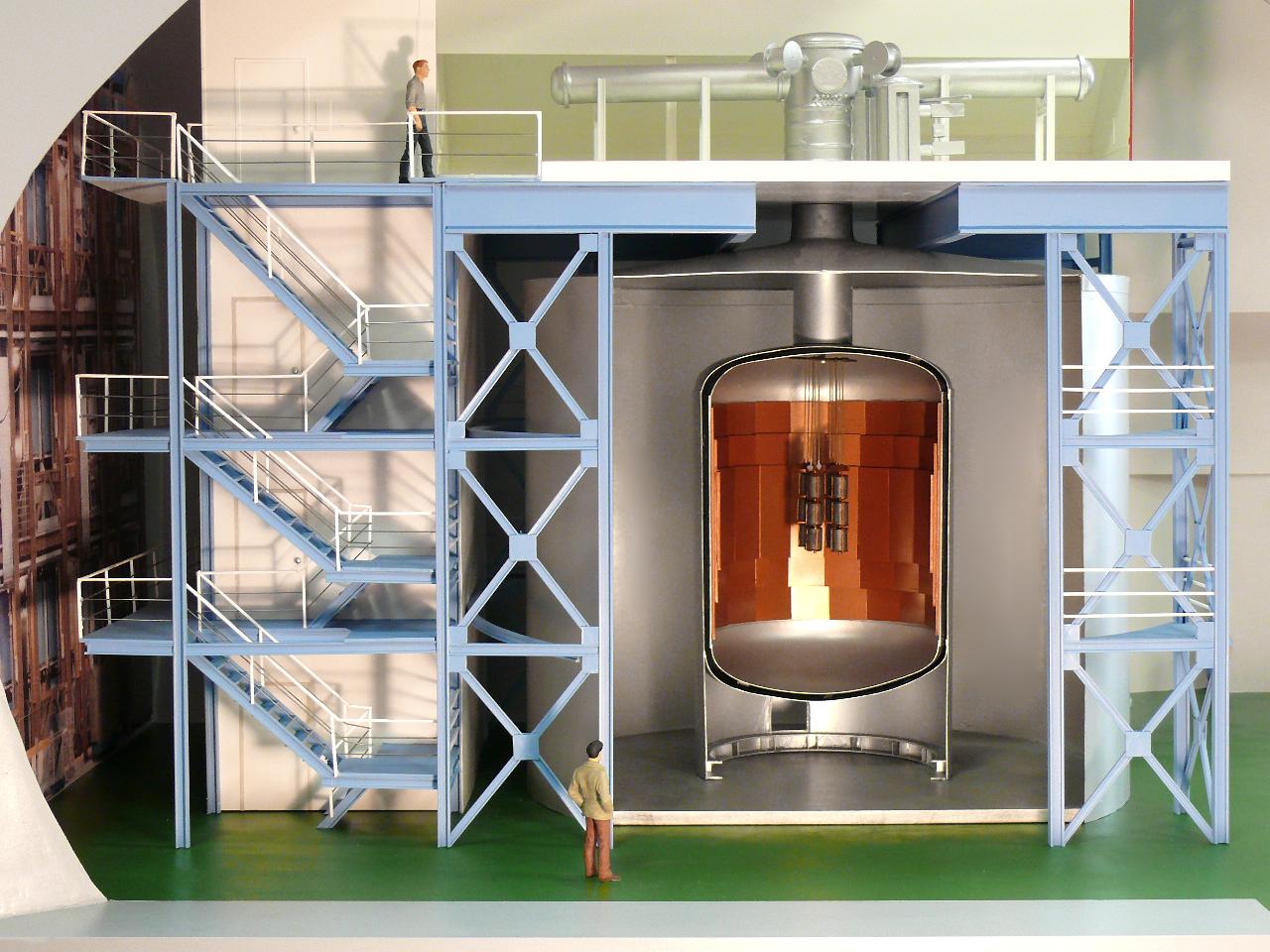}}
  \caption[Artist's view of \textsc{Gerda} experiment.]{Artist's view of the \textsc{Gerda} experiment (Figure from~\cite{Gerda_det}). } 
    \label{fig:Gerda_det}
\end{center}
\end{figure}
\textsc {Gerda} is currently taking data with phase I of their detector at the Laboratori Nazionali del Gran Sasso.  The experiment consists of an array of high-purity germanium detectors (HPGe) operated bare in liquid argon that acts as a cooling fluid and shielding from external backgrounds (see Figure~\ref{fig:Gerda_det}).  The array consists of 8 detectors enriched to 86\% in $^{76}$Ge and three natural germanium detectors for a total of about 25~kg.  The enriched detectors from the Heidelberg-Moscow~\cite{HDM} and \textsc{Igex}~\cite{igex} experiments.  The results discussed here are from data obtained between November 9, 2011 and March 21, 2012 with a live time of 125.9 days and an exposure of 5.04 kg$\cdot$yr.  The analysis has been carried out using six of the enriched germanium detectors.  The $2\nu\beta\beta$ half-life is determined to be $T_{1/2}^{2\nu} = 1.84^{+0.14}_{-0.10} \times 10^{21}$~yr~\cite{Gerda_2nu}.  Details of the analysis can be found in~\cite{Gerda_2nu}.  Data is still being taken for the $0\nu\beta\beta$ analysis and the ROI around $Q_{\beta\beta}$ is blinded for the analysis.

\section{Conclusion}

\begin{figure}
\begin{center}
  \resizebox{8cm}{!}{\includegraphics{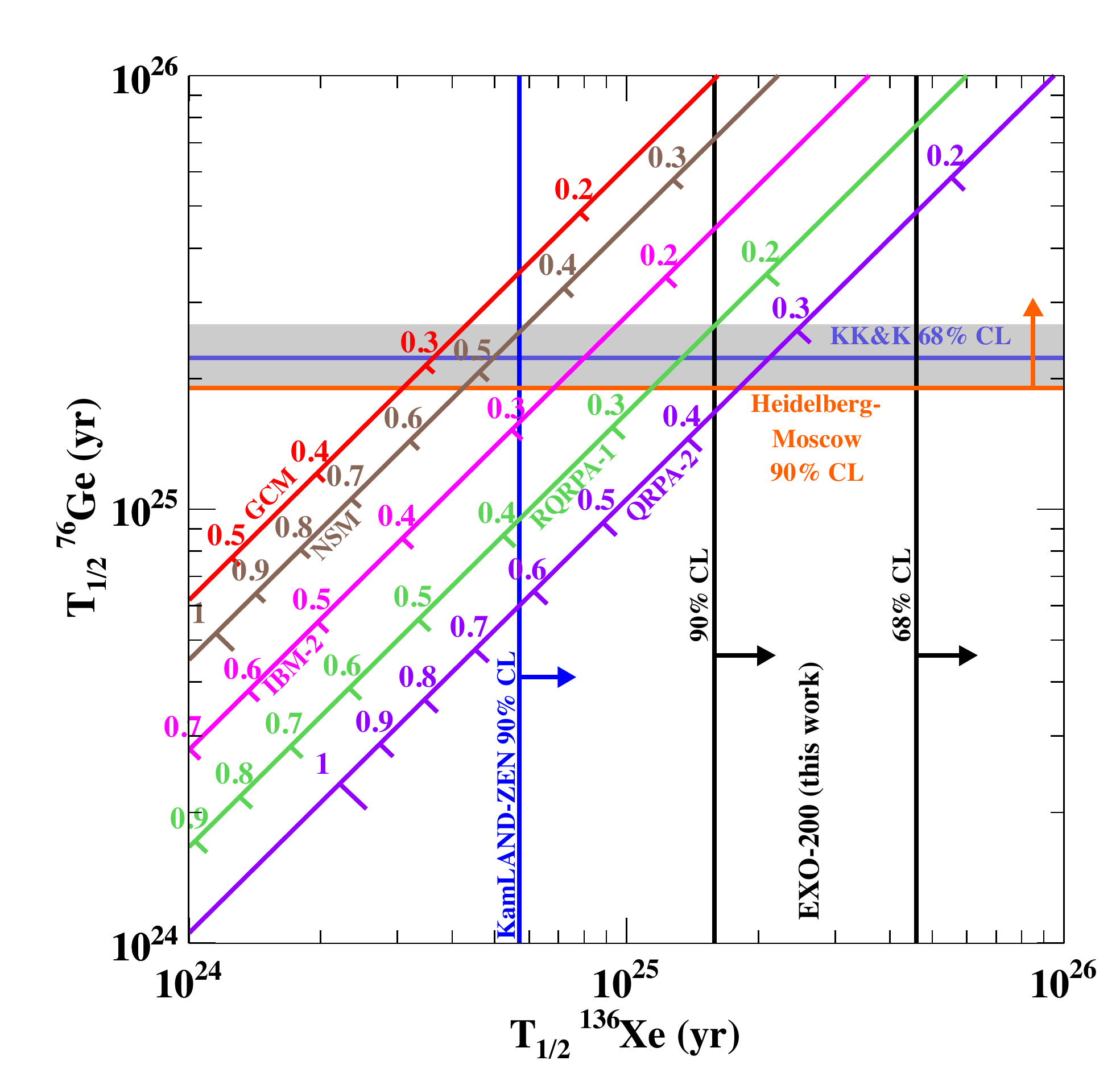}}
  \caption[Comparison between xenon and germanium double beta decay experimental results.]{Relation between the $T^{0\nu\beta\beta}_{1/2}$ in $^{76}$Ge and $^{136}$Xe for different matrix element calculations (GCM~\cite{GCM}, NSM~\cite{NSM}, IBM-2~\cite{IBM-2}, RQRPA-1~\cite{QRPA1} and QRPA-2 \cite{QRPA2}).   For each matrix element $\langle m_{\beta\beta} \rangle$ is also shown (eV).  The claim~\cite{Klapdor} is represented by the grey band, along with the best limit for $^{76}$Ge~\cite{heimo}.  The result reported here is shown along with that from~\cite{KLZ_2nu}.} 
    \label{fig:comparison}
\end{center}
\end{figure}

The  EXO-200 and KamLAND-Zen results are consistent with one another and nearly rule out the current claim~\cite{Klapdor}.  The comparison between germanium and xenon experiments for various nuclear matrix elements can be seen in Figure~\ref{fig:comparison}.  EXO-200 and KamLAND-Zen continue to take data to search for the $0\nu\beta\beta$ decay with a reach to an effective Majorana neutrino mass around 100 meV.  When \textsc{Gerda} phase I finishes running with its full exposure, it will provide a definitive test of the claim using the same isotope as~\cite{Klapdor}.

The next generation experiments: nEXO, KamLAND-Zen, CUORE, \textsc{Majorana} and \textsc{Gerda} are being designed to reach the inverted mass hierarchy ($\sim$ 50~meV) for neutrinos and improve the sensitivity to the $0\nu\beta\beta$ half-life by a factor of 100 or more.

\end{document}